\documentclass[review]{elsarticle}

\usepackage{amssymb}
\usepackage{array}
\usepackage{graphicx}
\usepackage{xcolor}
\usepackage{mathtools}
\usepackage{tensor}
\usepackage{graphicx}
\usepackage{amsmath}
\usepackage{amssymb}
\usepackage{mathrsfs}
\usepackage{color}
\usepackage{cancel}
\usepackage{tikz}
\usepackage{amsmath}
\usepackage{leftidx}

\def\beq{\begin{equation}}
\def\eeq{\end{equation}}
\def\bea{\begin{eqnarray}}
\def\eea{\end{eqnarray}}

\begin{document}

\begin{frontmatter}

\title{ Maximum rate of entropy emission}



\author{Behrouz Mirza}
\ead[url]{b.mirza@iut.ac.ir}
\author{Zahra Mirzaiyan} 
\ead[url]{z.mirzaiyan@ph.iut.ac.ir}

\author{Hamideh Nadi}
\ead[url]{h.nadi@ph.iut.ac.ir}
\address{Department of Physics,
Isfahan University of Technology, Isfahan 84156-83111, Iran}

\begin{abstract}
 It is shown that adding hair like electric charge or angular momentum to the black hole  decreases the amount of entropy emission. This motivates us to study the emission rate of entropy from black holes and conjecture a maximum limit (upper bound) on the rate of local entropy emission ($\dot{S}$) for thermal systems in four dimensional space time and argue that this upper bound is $\dot{S}\simeq k_{B} \sqrt{\frac{c^5}{\hbar G}}$. Also by considering R\`{e}nyi entropy, it is shown that Bekenstein-Hawking entropy leads to a maximum limit for the rate of entropy emission. We also suggest an upper bound on the surface gravity of the black holes which is called Planck surface gravity. Finally we obtain a relation between maximum rate of entropy emission, Planck surface gravity and Planck temperature of black holes.
\end{abstract}


\end{frontmatter}

 \section{Introduction}

The upper and lower bounds on physical quantities are usually very important and lead to interesting results. The most profound example is the upper limit on velocity which special relativity is based on. Lower bound on temperature which is now accepted as universal and embedded into the more general framework of statistical mechanics is another example. Christodoulou's idea of irreducible mass and Hawking area increase theorem are also other examples for bounded quantities in physics.\\
Our main purpose in this article is to add some important  bounded quantities to the list. We  obtain an upper bound on the force in GR and the rate of entropy emission ($\dot{S}$) for thermal systems in four dimensional space time and give an argument to show that this upper bound is $\dot{S}\simeq \sqrt{\frac{c^5}{\hbar G}}$. Surprisingly our result on proposing the maximum limit for the force is consistent with \cite{Gibbons2002,Gibbons2014}. It was proposed by Gibbons that value of a physical force in general relativity can not be more than $F_{max}=\frac{c^4}{4G}$. Here $c$ is the velocity of light and $G$ is the Newton constant of gravity. Numerically the maximum force is about $F_{max}\approx 3.25 \times 10^{43}$. Our result on maximum rate for entropy emission is quit novel. \\
 The conjecture for maximum force or tension in GR immediately  leads to an upper limit on the power or so-called Dyson Luminosity \cite{Dyson} as $P_{max}=\frac{c^5}{4G}$. Recently Gibbons  introduced a new conjecture that the dimensionless Schuster-Wilson-Blackett number, $c {\mu}/JG^{\frac{1}{2}}$, where $\mu$ is the magnetic moment and $J$ is the angular momentum, is bounded above by a number of order unity \cite{Gibbons2017}. In the same paper it is verified that such a bound holds for charged rotating black holes in those theories for
which exact solutions are available, including the Einstein-Maxwell theory, Kaluza-Klein theory, the Kerr-Sen black hole, and the so-called STU family of charged rotating supergravity black holes. All these efforts are looking for new fundamental bounds on physical quantities.\\\\
We find the rate of entropy emission for charged (RN) and rotating (Kerr) black holes with small electric charge and angular momentum. It is shown that adding any classic  hair to the black hole metric decreases the rate of entropy emission non-trivially. So we find that the rate of entropy emission is the most for a Schwarzschild black hole. This motivates us to ask if there is a maximum amount for the rate of entropy emission for a Schwarzschild black hole as a thermal system in nature. To answer this question we follow Page's work on the lifetime of black holes.
In $1976$ Don Page published his paper about black hole's radiation and evaporation time\cite{page}. By a numerical investigation \cite{toko1,toko2}, it was supposed that a black hole with mass $M\gg 10^{17}$ which emits massless particles including, $81  \% $ neutrinos, $17  \%$ photons and $2  \%$ gravitons should emit a total power of $2\times10^{-4} {\hbar}{c^{6}}{G^{-2}}{M^{-2}}$ . In \cite{page} the time of evaporation for a black hole is calculated numerically. We use Page's method to find the time a black hole evaporates (Black hole's lifetime) and sends information to the outside. We present a conjecture of the maximum rate of entropy emission. We use Planck's constants as $ m_{p}=(\frac{{\hbar}{c}}{G})^{\frac{1}{2}}=2.18\times 10^{-5} g$,\ \ $l_{p}=(\frac{{\hbar}{G}}{c^{3}})^{\frac{1}{2}}=1.62\times 10^{-33} cm$ and\ \ $t_{p}=(\frac{{\hbar}{G}}{c^{5}})^{\frac{1}{2}}=5.39\times 10^{-44} sec$. We also write the first law of thermodynamics and show that taking into account our conjecture on maximum tension, as well as our new conjecture leads to accept the familiar temperature $T_{\text{Planck}}$ as a maximum temperature for a black hole. We also propose an upper limit for the surface gravity of a Schwarzschild black hole.\\

It is known that the rate of emission of entropy from a Schwarzschild black hole is exactly the same as  a one dimensional quantum channel which means Schwarzschild black holes behave effectively as $(1 + 1)$-dimensional entropy emitters \cite{beke}. The case Reissner-Nordstr\"{o}m is also investigated analytically in \cite{Hod}. For recent works on black hole's quantum channel see \cite{Mirza2014}.  In this work our main purpose is to  find the maximum rate of entropy emitted from a single quantum channel for Schwarzschild black holes.

Beyond  GR, in Quantum Information Theory (QIT) there are some serious efforts to put bounds on the amount of information which a detector can gain. In \cite{ULC} a universal upper bound on the capacity of any communication channel between two distinct system is calculated. The mutual information transported between two systems is at most of order of $E {\Delta} t / \hbar$, where $E$ is the energy of the signal and ${\Delta} t$ is the amount of time which the detector operates.\\

The outline of paper is as follows: In Section II the rate of entropy emission for a Schwarzschild black hole is obtained. We find the power radiation and the rate of entropy emission for charged ($Q\ll M$) and rotating ($J  \ll M$)  black holes in Section III. In Section IV an upper limit for the rate of entropy emission is conjectured. In Section IV, We show that how the rate of entropy emission modifies considering R\'{e}nyi entropy rather than Bekenstein-Hawking definition. A maximum limit for the temperature for black holes is obtained in Section V by using the first law of thermodynamics and it is shown that this upper bound is the Planck temperature.  Also an upper limit is proposed for the surface gravity of black holes in Section VI. In Section VII a meaningful relation is founded between the maximum rate of entropy emission rate and the maximum information a black hole send to the surrounding environment. Section VIII is devoted to the conclusions. Also in Appendix we find how the rate of entropy emission and the force would be modified due to GUP (Generalized Uncertainty Principle).  At the end of the paper we list some bounded quantities in Physics.\\


\section{ Rate of entropy radiation for a Schwarzschild black hole}\label{S6}
Don Page has found the time of evaporation of a Schwarzschild black hole in 1976 \cite{page}. Here with the same method we find the power radiation and also find the rate of entropy emission for Schwarzschild black holes, assuming that these black holes emit only photons during the evaporation process. 
 Consider the Schwarzschild black hole metric in four dimensional space-time as

\begin{equation}\label{SCHW4}
ds^{2}=-(1-\frac{2GM}{c^{2} r})dt^{2}+\frac{1}{(1-\frac{2GM}{c^{2} r})} dr^{2}+r^{2} d{\theta}^{2}+ r^{2} sin^{2}{\theta} \ d{\phi}^{2}.
\end{equation}
The event horizon radius ($r_{h}$), the area of the event horizon ($A$) and the temperature ($T$) of the black hole are defined as follows 

\begin{equation}\label{SCHWRva T}
r_{h}=\frac{2GM}{c^{2}}, \ \ A=4 \pi r_{h}^{2}, \ \ T=\frac{{\hbar}{c^{3}}}{8\pi G k_{B} M}. \ 
\end{equation}

Due to Stephan-Boltzmann law the power radiated from a black body would be

\begin{equation}\label{Pof SCHW}
P={\sigma}_{4} A \ T^{4}\propto \frac{\alpha}{M^{2}},
\end{equation}
where, 

\begin{equation}\label{sigma va alpha}
{\sigma}_{4}=\frac{{\pi}^{2} {{k}_{B}}^{4}}{60{{\hbar}^{3}}{c^{2}}}, \ \ \ \alpha=\tilde{\alpha} \frac{{\hbar}{c^{6}}}{G^{2}},
\end{equation}
and $\tilde{\alpha}\simeq 2\times 10^{-5} $ is a numerical coefficient.\\

Black hole entropy is proportional to the area of the event horizon \cite{beken1973}, 

\begin{equation}\label{entSCh}
S=\frac{k_{B}  A}{4 l_{p} ^{2}}=\frac{k_{B} c^{3} A}{4 \hbar G}=\frac{\pi k_{B} c^{3}  r_{h} ^2}{ \hbar G}=\frac{4 \pi k_{B} G M^2}{\hbar c}.
\end{equation}

Using Eq.(\ref{Pof SCHW}), total power radiated from the black hole is

\begin{equation}\label{Pin m}
P=-\frac{dE}{dt}=-{c^{2}}{\frac{dM}{dt}}=\frac{\alpha}{M^{2}}.
\end{equation}
We have assumed a black hole which is evaporating and losing its mass as well as radiating energy to its outside surrounding and therefore, we expect a minus sign in Eq. (\ref{Pin m}).
Using Eqs. (\ref{entSCh}) and (\ref{Pin m}), we find the entropy emission rate as follows:
\begin{equation}\label{Sdot}
\dot{S}=\frac{dS}{dt}={\frac{8\pi k_{B} G}{\hbar c}}{M \dot{M}}=-\frac{8\pi k_{B} c^3  }{ G} \frac{\tilde{\alpha}}{M}.
\end{equation}\\

Note that if we work in natural unites where $k_B=c=G=\bar{h}=1$, the rate of entropy emission would be 

\begin{equation}\label{snew}
\dot{S}\propto -\frac{8}{M}.
\end{equation}

The above expression for the rate of entropy emission for Schwarzschild black holes is consistent with the rate of entropy emitted from a quantum channel for Schwarzschild black holes \cite{beke}.\\

In the next Section we find the the rate of entropy emission for the charged and rotating metric. We are interested to know how Eq. (\ref{Sdot}) will be changed by adding hairs to the black hole.

\section{ Rate of entropy emission for charged (RN) and rotating black holes (Kerr) in four dimensional space-time}

 In this Section we are interested to know how Eqs. (\ref{Pin m}) and (\ref{Sdot}) will change by adding an infinitesimal amount of charge ($Q\ll M$) and angular momentum parameter ($J \ll M$), where $Q$  would be an electrical charge and $a$ is the rotation parameter as $a=\frac{J}{M}$. \\

\textbf{I: RN metric with charge $Q\ll M$} \\

Consider RN metric in four dimensions as

\begin{eqnarray}
ds^2 =f(r) dt^2+\frac{dr^2}{f(r)}+r^2 d{\Omega}^2,
\end{eqnarray}
where,

\begin{eqnarray}\label{f(r)}
f(r)= 1-\frac{2M}{r}+\frac{Q^2}{r^2}.
\end{eqnarray}

To find the event horizon radius it is sufficient to solve the equation $f(r_{+})=0$,
so we have 
\begin{eqnarray}\label{EHR}
r_{+}=M+\sqrt{M^2 -Q^2}.
\end{eqnarray}

The area of the event horizon as well as the temperature reads as 

\begin{eqnarray}\label{AvaT}
A&=&4\pi r_{+}^2=4 \pi (M+\sqrt{M^2 -Q^2})^2, \nonumber\\
T&=&\frac{f^\prime (r_{+})}{4 \pi}=\frac{1}{4\pi r_{+}} (1-\frac{Q^2}{r_{+}^2})\nonumber\\
&=&\frac{1}{4\pi }(\frac{1}{M+\sqrt{M^2 -Q^2}}-\frac{Q^2}{(M+\sqrt{M^2 -Q^2})^3}).\nonumber\\
\end{eqnarray}

Now we impose the condition $Q\ll M$ or $\frac{Q}{M}\ll 1$. One can find the $r_+^d$ expansion at $Q\ll M$ as follows

\begin{eqnarray}\label{exp}
r_{+}^d\simeq (2M)^d-\frac{d}{4} 2^d M^{d-2} Q^2.\nonumber\\
\end{eqnarray}

Also using the above expansions for the radius, we use the expressions for area of the event horizon and temperature as follows
\begin{eqnarray}\label{entropy...dama}
A&=&4\pi r_{+}^2\simeq 4\pi (2M)^2 (1-\frac{Q^2}{2M^2}), \nonumber\\
T&\simeq &\frac{1}{8\pi M} (1-\frac{3Q^4}{16 M^4}).
\end{eqnarray}

Consequently $T^4$ would be
\begin{eqnarray}\label{T4}
T^4\simeq \frac{1}{(8\pi)^4 M^4} (1-\frac{3Q^4}{4M^4})\  \ (Q\ll M).
\end{eqnarray}
Now we can find the power which is radiated during the evaporation of RN black hole from Stephan-Boltzman formula as below

\begin{eqnarray}\label{p}
P=\sigma_{4} A T^4 \simeq \frac{2\sigma_{4}}{(8\pi)^3 M^2} (1-\frac{Q^2}{2M^2}+....).
\end{eqnarray}

The entropy of a charged black hole in the limit $Q\ll M $ would be

\begin{eqnarray}\label{s}
S=\frac{A}{4}=\pi (2M)^2 (1-\frac{Q^2}{2M^2})\simeq \pi (4M^2-2Q^2).
\end{eqnarray}

In order to find the rate of entropy emission, we need to find the power radiation ($P$) in terms of ($\dot{M}=\frac{dM}{dt}$). So we start with Smarr relation which defines the relation between mass and all other thermodynamic quantities. The Smarr relation for R-N black hole reads as
\begin{eqnarray}\label{smarrrn}
M=2TS+Q\varphi.
\end{eqnarray}
Also the electric potential is calculated and can be written as follows 

\begin{eqnarray}\label{potential}
\varphi=\frac{Q}{r_{+}}=\frac{Q}{2M}(1+\frac{Q^2}{4M^2})=\frac{Q}{2M}+\frac{Q^3}{8M^3}.
\end{eqnarray}
Using the Penrose definition, the energy which is radiated during the evaporation of a RN black hole with outer event horizon radius $r_+$  can be read as \cite{Eenery}
\begin{eqnarray}
E=M-\frac{Q^2}{2 r_+}.
\end{eqnarray}
Using Eq.(\ref{exp}) one can find the energy of a RN black hole with mass $M$ and electric charge $Q$ as
\begin{equation}
E=M-\frac{Q^2}{2 r_{+}}\simeq M-\frac{Q^2}{4M}+..... .
\end{equation}
So the power radiated in this limit would be
\begin{eqnarray}\label{power}
P=- \frac{dE}{dt}\simeq -\dot{M}+\frac{8M Q\dot{Q}-4Q^2 \dot{M}}{16M^2}.
\end{eqnarray}
We ignore the terms like $\frac{Q^{m}}{M^{n}}$ such that $n\geq m$. 

For Simplification we can assume 
\begin{eqnarray}\label{prn}
P \propto -\dot{M}+\frac{ Q\dot{Q}}{2M}+.... .
\end{eqnarray}

Using Eq. (\ref{s}), one could find the time derivative of the entropy, $\dot{S}$. The rate of entropy emission $\dot{S}$ is found to be 

\begin{eqnarray}\label{sdot}
\dot{S} \propto -8 (\frac{1}{M}-\frac{Q^2}{2 M^3})=-\frac{8}{M}+\frac{4Q^2}{M^3}.
\end{eqnarray}
Note that here the minus sign means entropy loss, so comparing with Eq.(\ref{snew}), It is found that so adding a small charge $Q$  decreases the rate of entropy emission by $\dot{S}_{\text{diff}}\simeq \frac{4Q^2}{M^3}$. It is important to note that decreasing or increasing the rate of entropy emission by adding any hair to black hole like electric charge is nontrivial and can not be known before deriving eq.(\ref{sdot}). \\

\textbf{ II: Kerr black hole with rotation parameter $J\ll M$ or equivalently $a=\frac{J}{M}\ll 1$,}\\

Thermodynamic quantities for the Kerr black hole read as,
\begin{eqnarray}\label{termoquan}
M&=&\frac{r_{+}^2 +a^2}{2 r_{+}},\\
S&=&\pi (r_{+}^2 +a^2)=\frac{A}{4},\\
J&=&\frac{a}{2 r_{+}} (r_{+}^2 +a^2),\\
\Omega &=& \frac{a}{(r_{+}^2 +a^2)},\\
T&=&\frac{1}{2\pi}[\frac{r_{+}}{(r_{+}^2 +a^2)}-\frac{1}{2r_{+}}].
\end{eqnarray}

We can find the radius of the event horizon by solving the following equation
\begin{eqnarray}\label{kerrradius}
r_{+}^2-2M r_{+} +a^2=0.
\end{eqnarray}
So the event horizon radius for Kerr black hole would be 
\begin{eqnarray}\label{kra}
r_{+}=M+\sqrt{M^2 -a^2}=M+\sqrt{M^2 -\frac{J^2}{M^2}}.
\end{eqnarray}

We impose the condition $J \ll M$. We would have the expansion for $r_+^d$ in the limit $J \ll M$ as follows

\begin{eqnarray}\label{exprkerr}
r_{+}^d \simeq (2M)^d-\frac{d}{4} 2^d M^{d-2} a^2.
\end{eqnarray}

The above expansion is so similar to the charged case just by replacing $Q$ with $a$.\\
The area of the event horizon and the temperature (which are the main ingredients to find $P$) may read as,
\begin{eqnarray}\label{kerrtna}
&&A=4\pi (r_{+}^2 +a^2) \simeq 4\pi(4M^2 -a^2)\simeq 4\pi (2M)^2 (1-\frac{a^2}{4M^2}),\\
&&T=\frac{1}{2\pi}[\frac{r_{+}}{(r_{+}^2 +a^2)}-\frac{1}{2r_{+}}]\simeq \frac{1}{8\pi M}(1-\frac{a^2}{4M^2}).
\end{eqnarray}

Consequently, $T^4$ can be read as

\begin{eqnarray}\label{T4kerr}
T^4 \simeq\frac{1}{(8\pi)^4 M^4}(1-\frac{a^2}{M^2}).
\end{eqnarray}

Now we can find power radiated during the evaporation process as,
\begin{eqnarray}\label{powerkerr}
P={\sigma}^4 A T^4 \simeq \frac{2{\sigma}^4}{(8\pi)^3 M^2} (1-\frac{5a^2}{4M^2}+...).
\end{eqnarray}

Similar to RN case we should now write the Smarr formula for the rotating black hole as,
\begin{eqnarray}\label{smarrkerr}
M=2TS+J \Omega,
\end{eqnarray}

\begin{eqnarray}\label{omega}
&&J=\frac{a}{2 r_{+}}(r_{+}^2 +a^2),\\
&&\Omega = \frac{a}{(r_{+}^2 +a^2)},\\
&&J \Omega = \frac{a^2}{2 r_{+}}=\frac{a^2}{4M}(1+\frac{a^2}{4M^2})\simeq\frac{a^2}{4M}.
\end{eqnarray}

So easily similar to our previous discussion on charged case,
\begin{eqnarray}
2TS=M-J\Omega \simeq M-\frac{a^2}{4M}.
\end{eqnarray}
The energy which is radiated
during the evaporation of a Kerr black hole has been investigated in \cite{Eenery,Eenery3}. The Penrose definition which is helpful to find the energy of a RN black hole during the evaporation, has not succeeded to deal with the Kerr metric \cite{Eenery2}. But for our purpose it is sufficient for us to know 

\begin{eqnarray}\label{kerrenergy}
E=M-\beta \frac{a^2}{r_{+}},
\end{eqnarray}
where $\beta$ is a positive integer and $\beta<1$ but still undetermined in \cite{Eenery,Eenery3}.
And the power radiated by a Kerr black hole with a small rotation parameter would be 
\begin{eqnarray}\label{powerkerrsmarr}
P=-\frac{dE}{dt}\simeq - \dot{M}+\beta \frac{a\dot{a}}{M}.
\end{eqnarray}

On the other side we find $\dot{S}$ as,

\begin{eqnarray}\label{sdotkerr}
\dot{S}\propto -\frac{8}{M}+\frac{2a^2}{M^3}+... .
\end{eqnarray}

One can see that adding a small rotation parameter $a$ can decrease the rate of entropy emission. It is also interesting to show the above results beyond a perturbative expansion in charges ($J \ll M$ and $Q \ll M$) numerically. \\

\section{Modification of Rate of entropy emission considering R\'{e}nyi entropy}

In the presence of gravity, in particular black holes, Boltzamann-Gibbs theory is out of validity. Also one has to take into account the long-range type property of the relevant interaction and the non-extensive nature of Bekenstein-Hawking entropy of black holes. Therefore, the Boltzamann-Gibbs statistics is no longer the best possible choice for defining the entropy in gravity. Non-extensive approaches to black hole entropy have been investigated with various methods. Consequently, there exist certain extensions of the Boltzamann-Gibbs entropy formula, which seem to be better choices to describe systems with long-range type interactions. One such statistical entropy definition has been proposed by Tsallis \cite{talis}. In case of the non-extensive Tsallis entropy, it turns out to be the famous R\'{e}nyi formula \cite{renyi1,renyi2} defined as

\begin{eqnarray}\label{ren}
S_R=\frac{1}{\lambda} \text{ln} [1+\lambda  \ S_{\text{T}}],
\end{eqnarray}
where, $\lambda\in  \mathbb{R}$ is a constant parameter and for a given discrete set of probabilities $\{p_i\}$, $S_{\text{T}}$ is Tsallis entropy, defined as $S_{\text{T}}=\frac{1}{1-q} \Sigma_i (p_i^q-p_i)$. In the limit of $\lambda \rightarrow 0$, both the Tsallis and the R\'{e}nyi formulas reproduce the standard Boltzmann-Gibbs entropy. In fact R\'{e}nyi entropy was introduced in information theory to provide the most general definition of information measures that preserve the additivity for independent events. For more details see \cite{reni1, morad,reni2}

 In this section, we show how the rate of entropy emission would be changed considering R\'{e}nyi entropy as the entropy of Schwarzschild black hole.  The R\'{e}nyi entropy of a black hole can be computed as \cite{reni2}
\begin{eqnarray}\label{ren}
S_R=\frac{1}{\lambda} \text{ln} [1+\lambda  \ S_{\text{BH}}],
\end{eqnarray}
where, we assume that $\lambda$ is a positive parameter and $\lambda \ll1$.
 For the Schwarzschild solution, the R\'{e}nyi entropy reads as
 \begin{eqnarray}\label{ren}
S_R=\frac{1}{\lambda} \text{ln} [1+\lambda  \ \frac{4 \pi k_{B} G M^2}{\hbar c}].
\end{eqnarray}

Using Eqs. (\ref{entSCh}) and (\ref{Sdot}) the rate of R\'{e}nyi entropy is found to be
 
\begin{eqnarray}\label{ren}
\dot{S}_R&=&\frac{\dot{S}_{\text{BH}}}{1+\lambda \ S_{\text{BH}}}\nonumber\\
&=&-\frac{8\pi k_{B} c^3\tilde{\alpha}}{G }\frac{1}{M}+\lambda \frac{32 \pi^2 k_B^2 c^2 \tilde{\alpha}}{\hbar }M.
\end{eqnarray}

Comparing Eq. (\ref{ren}) with (\ref{Sdot}), the rate of entropy emission decreases assuming the R\'{e}nyi entropy. As a result, it seems that as $\lambda$ is a positive integer, Bekenstein-hawking definition of entropy leads to the maximum rate for entropy emission while considering R\'{e}nyi definition of entropy.

\section{Maximum rate of entropy radiation}\label{S6}
Following our results in previous sections, we now know that adding any hair like charge or angular momentum to the generic metric decreases the rate of entropy emission. So we guess there should be a maximum amount for the rate of entropy emission like the force in GR. We are interested to know that if there is a possibility to find an upper bound for the rate of entropy emission.
 Note that we keep the constants $G,c, \hbar, k_{B}$ in our calculation.\\

 Let us integrate from the both side of equation (\ref{Pin m}), where $m$ is the initial mass of the black hole and $\tau$ is considered to be the evaporation time of the black hole. Using the last equality in Eq.(\ref{Pin m}) we have 

\begin{equation}\label{zamant}
-{\int\limits_{m}^{\epsilon}} m^{2} dm=\frac{\alpha}{c^{2}} {\int\limits_{0}^{\tau}}dt,
\end{equation}
where, $\epsilon$ is the residual mass of the black hole in evaporation process.
Therefore we obtain
\begin{equation}\label{tauschw}
\tau=\frac{{(m^3-\epsilon^3)}{c^{2}}}{3\alpha}.
\end{equation}
As we see the evaporation time of the Schwarzschild black hole in 4-dimensional space-time is proportional to $m^{3}$.\\
 Here we propose a conjecture for maximum rate of entropy emission using previous calculations. Consider the Schwarzschild black hole metric Eq.(\ref{SCHW4}). If we set $\tau=t_{\text{planck}}$ in Eq.(\ref{tauschw}), we can find the minimum mass as
 
 \begin{equation}\label{minmass}
 m_{min}=(3\tilde{\alpha})^{\frac{1}{3}} m_{p}.
 \end{equation}
  Now using equation (\ref{Pof SCHW}), we can find the maximum power in a thermal system by substituting $ m $ with $m_{min}$.
  \begin{equation}\label{maxp}
  P_{\text{max}}={\frac{\tilde{\alpha}}{(3\tilde{\alpha})^{\frac{2}{3}}}}{\frac{c^{5}}{G}}.
  \end{equation}
 Using eq.(\ref{maxp}) we can also find he maximum force in GR by
  \begin{eqnarray}\label{maxf}
  P_{\text{max}}=F_{\text{max}} . c\ \ \Rightarrow F_{\text{max}}={\frac{\tilde{\alpha}}{(3\tilde{\alpha})^{\frac{2}{3}}}}\frac{c^4}{G}.
  \end{eqnarray}
  
  It is interesting that our result for the maximum force and power in a thermal system, eqs. (\ref{maxp}) and (\ref{maxf}), are completely consistent with Gibbons maximum radiation power. Some years ago it was suggested by Gibbons that in general relativity there should be a maximum value to any physically attainable force and consequently he proposed a maximum value for the radiation power \cite{Gibbons2002,Gibbons2014}. \\

To find the maximum entropy emission rate we set $m=m_{min}$ in the eq. (\ref{Sdot}) as

\begin{equation}\label{maxs}
{\dot{S}}_{\text{max}}=-\beta  k_{B} \sqrt{\frac{c^5}{\hbar G}},
\end{equation}
where, $\beta=\frac{8\pi \tilde{\alpha}}{(3\tilde{\alpha})^{\frac{1}{3}}}\simeq 1.25\times10^{-2}$. Note that the minus sign shows that our black hole loses its entropy (information as thermal radiation) as a thermal system.\\

\textbf{We propose that the upper bound on the rate of local entropy loss for a thermal system is\\ ${\dot{S}}_{\text{max}} \simeq k_{B} \sqrt{\frac{c^5}{\hbar G}}\simeq 1.137\times 10^{20} kg/s^2 K $ }.\\

Note that $k_{B}=1.38 \times10^{-23} m^2 kg/s^2 K$, $\hbar= 6.62 \times 10^{-34}  m^2 kg/s$, $c=3\times 10^ {8} m/s$ and $G=6.67\times 10^{-11} m^3 /kg s^2$.

It is worthy to note that here we obtain an upper bound on the rate of local entropy emission, otherwise we could in general make our system bigger and bigger and so we would never have such an upper bound because we can make the radius of the shell as large as we like and from far away this shell looks like an isolated system with arbitrarily large power. Specifically by local entropy emission, we mean emission from a source which its entropy is proportional to its surface. So an upper bound on the  tension, radiated power and rate of entropy emission would be meaningless. Therefore the notion of rate of local entropy emission is important.\\
 We started with a Schwarzschild black hole with Planck mass ($m_{p}$) and consequently with radius ($l_{p}$). We call it a planckian black hole. We know that in this scale usual physics fail to describe what really happens to this black hole. note that we suppose that the evaporation process takes place from the initial mass and radius of black hole continuously until the black hole reaches to planckian size. It is interesting that if we calculate the black hole wave length $\lambda_{p}$, it will be of order $l_{p}$.
It means that maybe in the last step the black hole emits a quantum gravitational wave with the wavelength $l_{p}$ and the evaporation process ends.\\

\section{First Law of thermodynamics, Planck temperature}\label{S3}
In this Section, using our earlier arguments we obtain an upper limit on the temperature of thermal systems.\\
For perturbations of stationary black holes, the change of energy is related to change of area or entropy as 
\begin{equation}\label{1st law}
dE= TdS,
\end{equation}
where $dE$ is the change of energy which can be stated as the amount of work done by the system $dE=F.dx$ and $T$ is the temperature of the black hole. So the entropic force $F$ acting on the test particle is given by
\begin{equation}\label{Flaw}
F\frac{dx}{dt}=T \frac{dS}{dt}.
\end{equation}
We would like to know what  can be obtained if we maximize the L.H.S by substituting the maximum force, equation (\ref{maxf}) as $F_{\text{max}}=\frac{c^4}{G}$ and $\frac{dx}{dt}=c$ and devide it by ${\dot{S}}_{\text{max}} \simeq  k_{B} \sqrt{\frac{c^5}{\hbar G}}$ which was found in our earlier arguments.

\begin{equation}\label{how T}
F_{\text{max}}.c=T^\prime \dot{S} _{\text{max}}.
\end{equation}

So the obtained temperature can be written as below:
\begin{equation}\label{maxT}
T^\prime \simeq \frac{1}{k_{B}}\sqrt{ \frac{\hbar c^5}{G}}\simeq T_{\text{Planck}}.
\end{equation}
Therefore what is now obvious is that Gibbons maximum tension as well as our new conjecture on the rate of emitted entropy lead to the Planck temperature which is conjectured to be  the maximum possible temperature for black holes or thermal systems in general.

\section{Surface gravity and tension}\label{S4}
 In the previous Section, we found that considering the maximum tension as well as maximum rate of emitted entropy leads us to accepting the Planck temperature as the maximum possible temperature. In this Section, we use the relation between temperature and surface gravity of black holes and find an upper limit for the surface gravity.\\
The surface gravity is the force required that an observer at infinity to hold a particle (with unit mass) stationary at the event horizon.
 The Hawking temperature of black holes is read as 
 \begin{equation}\label{BHTem}
 T_{H}=\frac{\hbar \kappa}{2\pi c k_{B}},
 \end{equation}
 where, $\kappa$ is the surface gravity. To find the maximum value for surface gravity, one has to maximize the L.H.S by $T_{H} \simeq T_{\text{Planck}}$.
 
 Easily the maximum value for surface gravity is obtained by
 \begin{equation}\label{maxk}
 {\kappa}_{\text{Planck}}\simeq \sqrt{\frac{c^7}{\hbar G}},
 \end{equation}
 which is called Planck surface gravity .According to the definition of surface gravity, the surface gravity of a static Killing horizon is the acceleration, as exerted at infinity, needed to keep an object at the horizon.
 In Fig.1 we summarize the relations between $\dot{S}_{\text{max}}$, $T_{\text{Planck}}$ and $ {\kappa}_{\text{Planck}}$.

\begin{figure} 
\centering
\rotatebox{0}{
\includegraphics[width=0.7\textwidth,height=0.27\textheight]{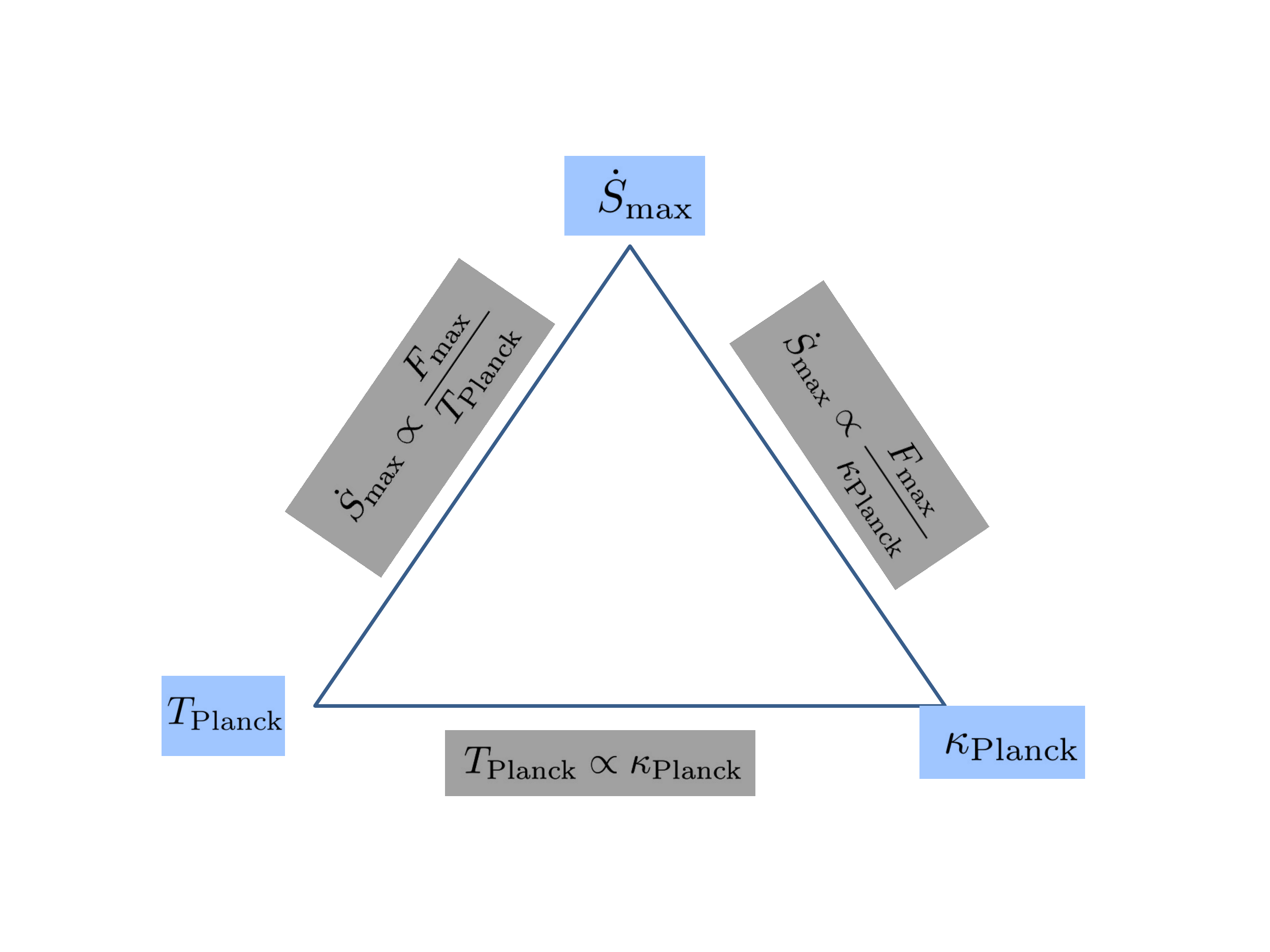}}\\
\caption{How maximum values relate to each other.
}\label{figure:RE}
\end{figure}

\section{Maximum rate for information emission }\label{S5}

In this Section, we obtain the maximum rate of information that a black hole can send to its surroundings which is related to our conjecture Eq.(\ref{maxs}).

Suppose Alice controls an arbitrarily bounded closed region of space with radius $R_{A}$ with some arbitrary content of matter and energy. Bob has surrounded Alice with the radius $R_{B}$ by some detectors. Alice would like to send a message to Bob who resides in a much vaster sphere. Note that: \\

 I. $R_{B}\gg R_{A}$\\
 
 II. There is no upper bound or restrictions on value of  either $R_{A}$ or $R_{B}$.\\
 
 III. Gravity doesn't have to be weak in Alice location (This is always doable by increasing $R_{A}$ and diluting system). Gravity is weak at $R_{B}$ as this is always the case in large radius in an asymptotically flat space-time.\\
 
 IV. Alice want to prepare a signal with energy $E$ to send it to Bob but the time for preparing such a signal is not important. She can get an arbitrarily early start.\\
 
 V. The time period during which Bob's detectors are operating is known to Alice.\\
 
 With all the above assumptions there is an upper bound on the amount of information Bob can get due to quantum effects?
  
 If we consider Alice and Bob as two distinct systems there is a communication channel between them.
 We follow this arrangement as we suppose Alice as an evaporating black hole since there is no limit on the strength of gravitational field for Alice.  Bob is the surrounding environment and the information (signals with energy $E$) is the entropy $\widetilde{S}$, which is emitted to the outside.
If we consider a black hole which is sending signals with maximum temperature $E=k_{B} T_{\text{Planck}}$ in an arbitrary amount of time, we would expect that $\frac{d\widetilde{S}_{\text{max}}}{dt}\leq \dot{S}_{\text{max}}$, where $\dot{S}_{\text{max}}$ is the upper bound on the rate of local entropy emission which is obtained as a new conjecture in Section II. We claim that $\dot{S}_{\text{max}}$ is the maximum rate of entropy emission so naturally we expect that any quantity which is related to the rate of entropy be smaller or equal to $\dot{S}_{\text{max}}$. It can be easily seen that this happens and
 
 \begin{equation}\label{ULCcon}
\frac{d\widetilde{S}_{\text{max}}}{dt}\leq \dot{S}_{\text{max}}\sim k_{B} \sqrt{\frac{c^5}{\hbar G}}.
 \end{equation}
 
As a result our conjecture shows the Maximum rate for information emission which a black hole can send to its surroundings.

\section{Conclusion}\label{S7}
In this paper we propose a conjecture for the maximum rate of entropy emission for a thermal system. We show that our result is consistent with Gibbon's conjecture for maximum tension in GR. We support our conjecture by showing that adding hairs like electric charge or angular momentum to the black hole metrics decreases the rate of entropy emission.  We also used our conjecture along with Gibbon's work and showed that accepting these maximum values leads to finding the Planck temperature as the maximum possible temperature. Based on the highest possible temperature, an upper bound for surface gravity is obtained and using the maximum surface gravity, we reobtained Gibbon's conjecture on the maximum force in GR. We also showed that the new conjecture on rate of entropy emission can be related to the maximum amount of information a black hole can send to the surrounding environment.\\

\begin{center}
\begin{tabular}{ | m{12em} | m{10em}|  } 
\hline
Quantity& Maximum Value  \\ 
\hline
Maximum Tension & $F_{max}\simeq\frac{c^{4}}{G}$ \\
 
\hline
Maximum Radiation Power & $P_{max}\simeq\frac{c^{5}}{G}$   \\ 

\hline
Maximum Rate of Entropy Emission & $\dot{S}_{max} \simeq k_{B} \sqrt{\frac{c^5}{\hbar G}}$ \\

\hline
Maximum Temperature & $T_{Planck}\simeq\frac{1}{k_{B}}\sqrt{ \frac{\hbar c^5}{G}}$ \\

\hline
Maximum Surface Gravity & $k_{Planck}\simeq \sqrt{\frac{c^7}{\hbar G}}$ \\

\hline
\end{tabular}
\end{center}


\end{document}